# Spectroscopic Observation and Modeling of Photonic Modes in $CeO_2$ Nanocubes

Yifan Wang, Shize Yang, and Peter A. Crozier*

**Abstract**

Photonic modes in dielectric nanostructures, e.g., wide gap semiconductor like $CeO_2$ (ceria), has potential for various applications such as light harvesting and information transmission. To fully understand the properties of such phenomenon in nanoscale, we applied electron energy-loss spectroscopy (EELS) in scanning transmission electron microscope (STEM) to detect such modes in a well-defined ceria nanocube. Through spectra and mapping, we demonstrated a geometrical difference of mode excitation. By comparing various spectra taken at different location relative to the cube, we also showed the transmission properties of the mode. To confirm our observation, we performed EELS simulation with finite-element dielectric calculations in COMSOL Multiphysics. We also revealed the origin of the modes through the calculation. We purposed a simple analytical model to estimate the energy of photonic modes as well. In all, this work gave a fine description of the photonic modes' properties in nanostructures, while demonstrating the advantage of EELS in characterizing optical phenomena in nanoscale.

# Introduction

Photonic modes, sometimes referred as cavity modes or waveguide modes, is the optical response from dielectric material with specific sizes and shapes. Photons travelling through the structure with specific wavelength will be trapped in the form of photonic modes. Such modes have been widely applied in fields such as telecommunication [1,2], laser generation [3,4] and sensors [5,6]. When the size of the structure has reached to nanoscale, i.e., the wavelengths of the modes are in the visible light region, it becomes possible to capture solar energy through this mechanism. Comparing to other visible light harvesting approaches, e.g., photovoltaics or plasmonic harvester, a significant advantage of photon capturing with photonic modes is that there is no loss from electronic excitation if the energy of the mode is below the bandgap. Moreover, the modes are straightforward to engineer since the energy of the modes are solely dependent on the geometry of the structure and the dielectric properties of the material. By coupling with metal nanoparticles, energy from photonic modes can be transferred to the catalytic active sites on metal. [7] In other words, such dielectric nanoparticle-metal system can act as a photocatalyst for reactions like water splitting [8].

To fully tap the potential of this mechanism for catalytic application, a deeper understanding of these modes' properties is needed. However, despite high energy resolution, traditional optical measurement methods, e.g., Raman spectroscopy [9] and infrared spectroscopy [10], usually suffers from low spatial resolution to around 1 micrometer. A way to achieve high spatial resolution is using high energy electron beam as the light source, which has a much shorter wavelength than photons. For a free electron beam, the electric field in frequency space are in the form of plane waves, giving rise to virtual photons. [11] Therefore, it can excite optical responses in materials nearby. An example of this is Smith-Purcell effect, in which the swift electron travels parallel by a diffraction grating and forms visible light with certain wavelength [12]. In the case of scanning transmission electron microscopy (STEM), high energy electron probe can act as a continuum light source with atomic resolution.

With the recent development on electron optics, especially aberration corrector [13,14], monochromator [15,16], and detectors [17,18], studying low energy features including plasmonic [19–21] and phononic [22–31] response at atomic resolution with adequate energy resolution (~10meV) through electron energy-loss spectroscopy (EELS) in STEM becomes feasible. Also, development of aloof beam technique and theory makes detection on thick objects that scatter electron beyond detection limit possible. [32] The high spatial resolution of STEM imaging, especially high angle annular dark-field (HAADF) imaging also grants the capability to visualize the spatial distribution of energy-loss events. [33]

Previous studies on photonic modes in silicon-based materials with different shapes supported by thin films has proved STEM EELS as a powerful tool to explore this phenomenon [34–36]. Also, since some of the system have good cylindrical symmetries, e.g., disks and ribbons, analytical calculations are performed. Numerical calculations on more mathematically complicated system, e.g., ellipses and triangle pillars, also gives a deeper understanding of the modes. Also, silicon-based materials have a bandgap in visible light region, therefore not suitable for light harvesting. Studies on stand-alone dielectric nanoparticles with wider bandgap, e.g., $CeO_2$, and $TiO_2$ has also demonstrate the existence of photonic modes. [37] However, due to the rather irregular geometry of the nanoparticles, there are limitations on interpretation of the spatial related properties of photonic modes.

In this paper, we performed monochromated STEM EELS on well-defined cerium oxide nano-cubes. Through EELS point spectra and mapping, we demonstrate different mode excitation due to the geometry difference, i.e., some modes be more preferentially excited when the beam is placed at specific positions. Besides, we also discuss the transmission properties of the photonic modes. To further understand such spatial difference in mode excitation, numerical finite element calculation was performed in the commercial software COMSOL Multiphysics to visualize the modes. A simple analytical model was used to calculate the energy of photonic modes in cubic nanoparticles with given dielectric functions.

**Methods**

Ceria nanocubes with predominantly (100) surfaces were synthesized using hydrothermal method [38]. Cerium nitrate hexahydrate ($Ce(NO_3)_3·6H_2O$) and sodium hydroxide (NaOH) were separately dissolved in deionized water. The solution was mixed and stirred on a magnetic stirrer for 30 min. The final molar concentration of NaOH was calculated to be 8 M. In order to get cubes with ideal size for photonic modes, the mixture was heat at 220°C for 24 hours and cooled down naturally. The precipitates, in form of powder, were dispersed in deionized water by sonicating for 20 min. The top layer of the suspension is used for TEM observation by drop-casting and baked on a copper grid with lacey carbon film.

Monochromated-EELS was performed with NION UltraSTEM 100 microscope equipped with aberration corrector and the state-of-the-art Dectris Ela detector [18]. The microscope was operated at 100 kV. The energy dispersion of the spectrometer was set to 5 meV per channel. Measured full width half maximum is 18 meV. The convergence and collection semi angle of the experiment were set to 19 mrad and 21.8 mrad, with a 2 mm EELS entrance aperture. To increase the signal to noise ratio (SNR), 100 spectra taken at the same position with 500 ms exposure time are integrated to form a point-and-shoot spectra. EELS mappings were also done with a dwell time of 100 ms per pixel to minimize the effect of sample drifting while maintaining an adequate SNR. The data was processed with Nion Swift and Gatan Digital Micrograph.

Since the experimental setting is complicated for analytical calculation, the numerical finite element method is applied here. The calculation is performed with the commercial software COMSOL Multiphysics, which has been proved effective within plasmonic loss [39,40] and vibrational loss region [41]. Based on the cerium oxide dielectric function [42], Maxwell equations can be solved numerically in frequency space. Electric field corresponding to different frequencies can be derived from the solution of Maxwell equation. Besides, energy loss probability can be evaluated through the following equation [43]:

$$\Gamma_{EELS}(\omega) = \frac{ve}{2\pi\hbar\omega}\int dz Re[E_z^{ind}(z,\omega)e^{-\frac{i\omega z}{v}}]$$

, where $v, e, \omega, E^{ind}$ are the velocity of electron, unit charge, frequency and induced electric field, and z direction is the propagating direction of electron beam respectively. To acquire the cavity eigenmodes and the corresponding electromagnetic field in a nano-cube, simulation is performed with the wave-optic module in COMSOL, which has been proved effective in calculations related to photonics with even more complicated structure, e.g., photonic crystal [44]. Through eigenfrequency searching, possible frequencies of cavity modes are calculated in a certain energy region (2-4eV).

**Result and Discussion**

A ceria nano-cube with a size of around 200nm in a [100] orientation is shown in the high-angle annular dark-field (HAADF) image of **Figure 1a**. One face of the cube is nearly clean. The clean surface minimizes the possible disturbance from smaller particles which are clustered on other surfaces. Monochromated EELS in the aloof configuration was performed since the sample was too thick for 100 keV electrons to penetrate and without being scattered beyond EELS collection angle [45]. To prevent transmission between part of the convergent beam and the nanoparticle, the impact parameter, *b*, which is the distance between probe and the particle surface, was set to 10 nm. **Figure 1b** shows spectra from two different points: the blue one, denoted as the edge setting, is near the top left edge of the cube which is parallel to the direction of the incident electron beam, and the red one, denoted as the face setting, is around the middle of the face of the cube. Four major peaks between 2 eV and the bandgap edge (~3.40eV), denoted as peak I(~2.45eV), II(~2.65eV), III(~3.10eV), and IV(~3.25eV), are presenting in the spectra. In the face center geometry, peaks II and III are the predominant peak, while peak IV is about half the intensity of peak III, and there is nearly no sign of peak I. For the edge setting, peaks I and IV have a relatively stronger signal comparing to face configuration, while there is a peak II$^*$ of low intensity that have energy very close to peak II, and peak III is covered by the tail of peak IV.

**Figure 2** presents the background-subtracted EELS intensity mapping of each major peaks across the cube. An EELS mapping is taken from the cube and surrounding area, and the background-subtracted intensity at each major peaks' (I, II, III, IV) energy is extracted as the pixel's value. The ceria cube is presented as a square of low signal at the right top of each figure. This is because in transmission, electrons will have higher probability of being scattered beyond the collection angle of the spectrometer than in the aloof configuration, thus causing a suppression of the overall intensity. For the same reason, in all four maps, the area beneath the cube has some irregular medium-low intensity contrast, which are corresponding to the small particles decorating the surface of the cube shown in the HAADF image (**figure 1a**). For peak II and III, the high intensity region formed an arc-shape area, and the brightest area is at the center of the arc, close to the middle of the (100) surface. Meanwhile, peak I and peak IV give higher intensity near the edge of the cube, especially the top left and bottom right. Also, peaks I and IV are more strongly excited around the edge of the cube while modes corresponding to peaks II and III are more strongly excited at the middle of the face.

Notice that the peaks' intensities in the map do not show perfect spatial symmetry, especially the mapping for peak IV. To understand the slight asymmetry, we acquired multiple spectra at different points along the side of the cube and extracted signals at several equivalent points in terms of the geometry of the cube. **Figure 3** shows two sets of 3 spectra at symmetrically equivalent points which shows unexpected differences in major peaks for both settings, matching result of the EELS mapping For face setting (**figure 3b**), spectra at f1 and f3 are giving qualitatively identical results as the face setting spectrum in **figure 1b**, while f2 shows an intensity decrease and peak broadening for both peak II and III. Similar results are observed in the edge settings (**figure 3c**): e1 and e3 are equivalent to the edge setting result in **figure 1b**, whereas peak I and especially, peak IV are suppressed when beam is placed at e2. Although two sets of points are geometrically equivalent to the cube itself, the surrounding environment of the cube, e.g., small cubes attached to the large cube, do not guarantee a perfect symmetry. There are two main elements that can possibly lead to disturbance of the mode: (1) small particles decorated on the surface (2) the huge cluster at the top right of the cube. According to the HAADF image, e1, f2 and f3 are the three settings that are closer to the 'unclean' part of the surface, i.e., amplifying the effects due to the small cubes on the surface. The conduction band signal is stronger and shows a steeper rise demonstrating that it is affected by the small cubes which are closer to the probe at point e1, f2 and f3. However, despite the fact that all of these three points collect signal from small cubes, only f2 shows damping of photonic peak intensity, which confirms that the asymmetry in the excitation of photonic modes is not caused by the small particles on the surface of the cube. Meanwhile, if we consider the possible boundary conditions of electromagnetic standing waves in the crystal, we can see that modes generated from e1, e3, f1 and f3 are predominately bounded by vacuum interfaces, whereas for e2 and f2, one of the boundaries is associated with large aggregates with dimensions larger than the cube (see Figure 1a). This allows the photonic mode to couple into the adjacent aggregates and changes the resonant frequencies. (For more information on coupling the reader is referred to Liu et al [37]). In other words, photons have higher probability propagating from vacuum to the cube, then to the large cluster. These observations demonstrate that the variations in spectra from photonic modes are caused by the large aggregates at the top right side of the cube.

To aid the understanding in the origin of the change of the intensity and broadening of the peaks, it is useful to look at the physical interpretation of the peak in a spectrum. A peak can be described by its mode energy ($E$), integrated intensity ($\int IdE$, where $I$ is the intensity), which is proportional to total number of photons generated, and characteristic width ($\Delta E$). By time-energy uncertainty relation [46]: $\Delta E \Delta t \simeq \hbar$, a broad peak (large $\Delta E$) means that the lifetime of the mode ($\Delta t$) will be short. Therefore, in terms of lifetime, the photonic modes decays much faster when the beam was placed at the point e2 and f2 than e1, e3 and f1, f3 according to **figure 3**. From the simple case of Fresnel law, when electromagnetic (EM) wave travels normal to the interface, the reflected coefficient ($R$) and transmitted coefficient ($T$) of two linear media are [47]:

$$R \equiv \left(\frac{n_1 - n_2}{n_1 + n_2}\right)^2 \quad T \equiv \frac{4n_1 n_2}{(n_1 + n_2)^2}, \quad R + T = 1,$$

where $n_1$ is the refractive index of the medium that the EM wave comes from, and $n_2$ is the refractive index of the adjacent medium. If the two refractive indices are close to each other, $R$ will be close to 0 while $T$ will be close to 1, which means most of the wave will be transmitted through the interface instead of reflecting back from the interface. For cavity modes, a larger $R$, will increase the probability that the photon will be reflected from the surface and trapped inside the nanocube. Thus, the standing wave it forms has a longer lifetime, which will be shown as a sharper peak in spectra. According to Maxwell Garnett effective medium theory, the average dielectric function of the large cluster (including vacuum) can be approximated as [48]:

$$\epsilon_{avg} = \epsilon_{vac}\left[1 + \frac{3f\left(\frac{\epsilon_{CeO_2} - \epsilon_{vac}}{\epsilon_{CeO_2} + 2\epsilon_{vac}}\right)}{1 - f\left(\frac{\epsilon_{CeO_2} - \epsilon_{vac}}{\epsilon_{CeO_2} + 2\epsilon_{vac}}\right)}\right]$$

where $\epsilon_{avg}, \epsilon_{vac}, \epsilon_{CeO_2}, f$ are the dielectric function of the cluster, vacuum and ceria, and the volume fraction of ceria respectively. From this equation we know that the average refractive index of the cluster will lie between the refractive index of vacuum and pure ceria. Thus, for waves travelling to the large cluster, i.e., modes generated by electron beam positioned at e2 and f2, they will decay faster due to higher probability of energy transmission to the clusters. When the beam is placed at e1 and e3, most energy will reflect thus forming an intense standing wave, giving a sharper peak in EELS (**figure 3c**). For the f1 and f3, although the right surface is not perfectly clean, the thickness of the particle layer is much smaller comparing to both the size of the cube and the size of the large cluster above the cube. Thus, instead of having the mode directly transmitted to large clusters, there is still a higher possibility for the wave trapped inside the cube.

To further understand the photonic modes in the cube, we performed a numerical simulation with COMSOL Multiphysics. The geometry for the simulation is shown in **figure 4a** and was similar to the experiment set-up with an incident electron energy of, 100 keV. We use a line current to simulate the electron beam, which is set 10 nm away from the cube and along z direction in our Cartesian coordinate system. The beam was placed at the same position as the experimental edge and face settings. The simulated electron energy loss spectra are plotted as **figure 4b**. The energy-loss spectra of two geometries show a very good qualitative match with the experimental spectra: the position of the valence loss edge is roughly the same, and the major four peaks and a minor peak II* and their relative intensity agrees in both settings of experimental and simulation spectra. The reason for slight difference between experiment and simulation might come from: (1) The dielectric data we use for simulation, which is measured from a film, may not 100% match with our actual cube. (2) The experimental spectra can be considered as the convolution of our simulation result and the ZLP, thus resulting in broader peaks that will cover minor features [49].

An intuitive way to visualize the modes is to plot the electric fields in the cube with each excitation geometry. We calculate the electric field in the cube for each peak energy. **Figure 5** shows the z components of electric field in x-y plane (normal to the travelling direction of beam). Here we select cross section through the cube that has highest intensity of the electric field. For low energy modes (peak I and II), the highest intensity cross section is in the middle of the cube, while low or even zero intensity is on the top and bottom of the cube. Whereas the higher energy modes give strong electric field at the 1/4 and 3/4 height of the cube, and has weak field at middle, top and bottom of the field. (See supplements for field at other heights.) Also, for the cross-section of same energy at different heights, while the shape of the field keeps nearly identical, the sign of the field has reversed. The behavior in z-direction indicates that a periodicity of electric field: a half period for lower energy peaks, and a full period for higher energy peaks. Apart from the variation in z-direction, the electric field also shows specific symmetry for two different settings. The electric field maps of two edge setting peaks show a symmetry along the diagonals, while for the face setting, the symmetry axis is the horizontal or vertical bisector line. The shape of the electric field also indicates the wave vector of the mode, especially for the edge modes: a wave pattern is formed in diagonal direction. In addition, from the other way around, an energy comparison can be done with the estimation of wavelength based on the electric field pattern. Peak I has a wavelength proportional to the diagonal of the square, thus the energy is lower than peak II, of which the wavelength is around the side of the cube. For peak III and II, despite they share similar electric field pattern in x-y plane, peak III has a shorter wavelength due to one half more period in z-direction, and therefore a higher energy. The wavelength of peak IV in x-y plane is two third of the diagonal, which is shorter than the side of the cube by a little amount. This matches the fact that the energy of peak IV is slightly higher than peak III.

Table 1a. Solution to Standing Wave Equation

| $m\ n\ l$ | 2 1 0 | 2 1 1 | 2 2 0 | 2 2 1 / 3 0 0 | 3 1 0 |
|---|---|---|---|---|---|
| $k^2$ | 5 | 6 | 8 | 9 | 10 |
| Energy (eV) | 2.68 | 2.87 | 3.15 | 3.25 | 3.36 |

Table 1b. Simulated Peak and Corresponding Energy

| Peak | I | II | II* | III | IV |
|---|---|---|---|---|---|
| Energy (eV) | 2.59 | 2.78 | 2.89 | 3.24 | 3.39 |
| Possible Mode ($m\ n\ l$) | 2 1 0 | 2 1 1 | 2 2 0 | 2 2 1 | 3 1 0 |

To understand these modes, we apply the basic standing wave model here [50]:

$$\omega_{mnl} = \frac{\pi}{\sqrt{\mu\epsilon(\omega)}} \left(\frac{m^2}{a^2} + \frac{n^2}{b^2} + \frac{l^2}{c^2}\right)^{1/2} = \frac{\pi}{\sqrt{\mu\epsilon(\omega)}} \frac{k}{a}, \ where\ k = \sqrt{m^2 + n^2 + l^2}$$

where $\omega, \mu, \epsilon, a, b, c$ are the frequency of the mode, the permeability (which we consider as a constant), the dielectric function in frequency, and three dimensions of the cube respectively. $m, n, l$ are non-negative integers, which stands for the wave mode for each direction. Since it is a cube, $a, b, c$ are the same and $a = 200nm$. In this case, the solution to this equation is given in **table 1a**. To confirm the result of theoretical model, we perform an eigenfrequency search using finite element model in COMSOL wave optic module. The result has a good agreement. (See the detail in Supplementary) By comparing the peak energy generated by simulated electron beam (**table 1b**) and the photonic eigenmodes, there is a roughly match between the energy of the theoretical solution and simulation. The reason for not perfect match is because, in this model and COMSOL wave optic calculation, we are assuming the cube as perfect boundaries, i.e., all the waves will be scattered beyond the boundary, which is not true in both experiments and EELS simulations.

**Conclusion**

We observed photonic modes in a well-defined ceria nano cube using STEM EELS point spectra and mapping. By placing the beam at different positions around the cube, we showed the existence of different modes and the variation of the intensity with different beam positions, i.e., edge sites and face sites. We utilized the slight asymmetry of the surrounding environment of nano cube to demonstrate the transmission properties of photonic modes. Photonic modes will be largely suppressed when the cube has large clusters that have a size comparable with the cube itself decorated on. In other words, energy will be transmitted to those clusters rather than trapped in the cube. EELS simulations through numerical calculation in COMSOL Multiphysics qualitatively agree with experiments. The calculation also provides electric field distribution in the cube. The field shows a standing wave pattern with wave vectors corresponding to different beam settings, i.e., a diagonal wave vector corresponds to edge settings, and a wave vector parallel to the side of the cube is generated from the face settings. Besides, optical calculation gives eigenmodes of the cube, which agrees with the numerical calculation. A simple equation is purposed here to calculate photonic modes in nano cubes given size and dielectric function, and the result has a good match with the numerical simulation. This work will provide guidance for engineering the energy of photonic modes, especially when the size control of dielectric nano cube can be very precise.

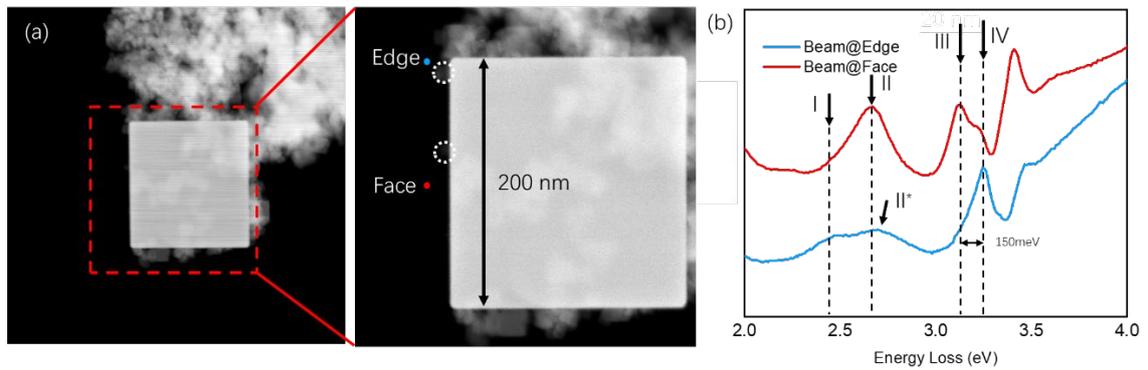

**Figure 1** (a) STEM image of the cube and the position of the two points where the spectra were taken. The left face of the cube is nearly clean except two tiny cubes, denoted by dotted circle. Both points are 10 nm away from the cube. (b) The background-subtracted energy-loss spectra at two points. The energy of the four peaks is indicated by arrows.

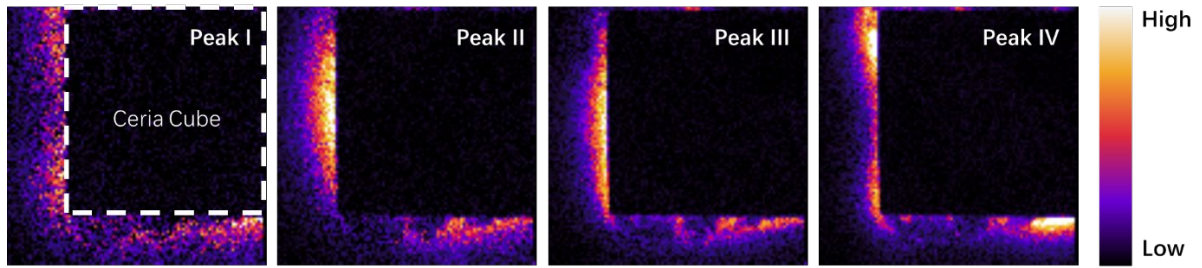

**Figure 2** The EELS intensity mapping of the cube and surrounding region. The white dotted box indicates the cube. The peak intensity is represented by the blackbody temperature of the color. Cold color (purple → black) suggests a low intensity while warm color (orange → white) indicates a high intensity. Brightness and contrast of the four images are adjusted independently to stress the effect.

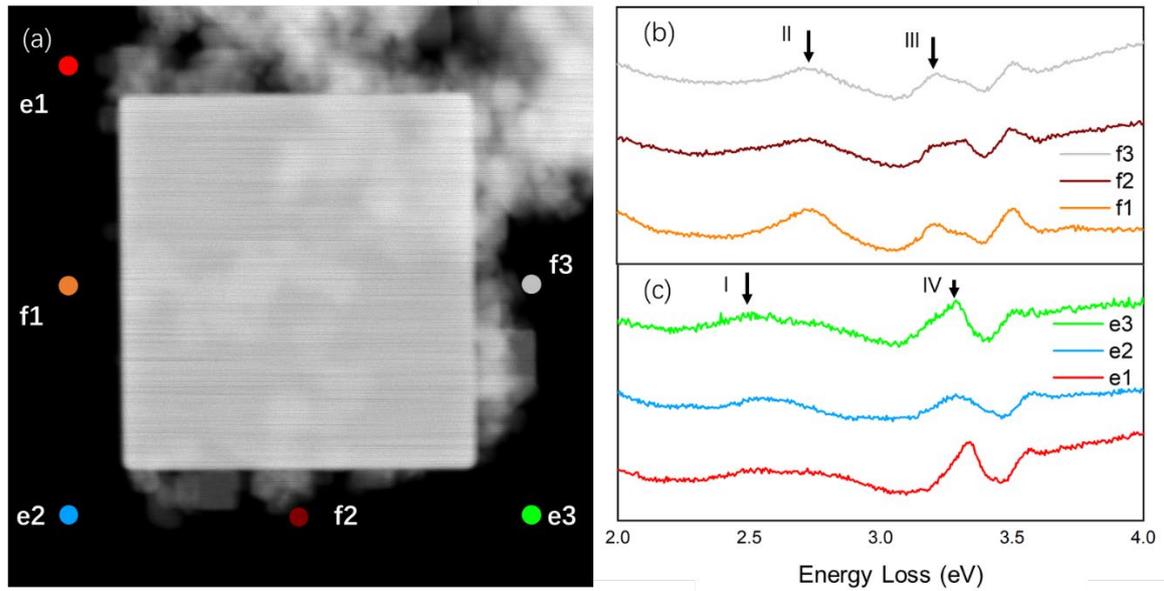

**Figure 3** HAADF image and EELS point-scans extracted from EELS mapping at 2 set of geometrically equivalent points. (a) The 6 points are showed in the HAADF image. The edge settings are denoted by **e** and face settings are denoted by **f**. (b-c) Two set of spectra shows the energy loss spectra corresponding to each point. (b) presents the face settings and (c) presents the edge settings. All the spectra are normalized by zero-loss peak and plotted separately.

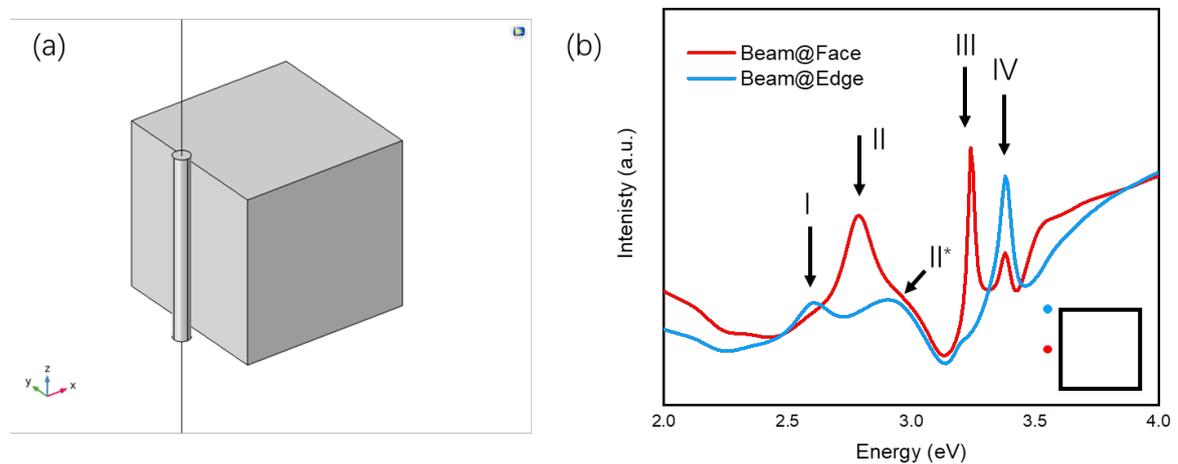

**Figure 4** The simulation setting and the simulated energy loss spectra. (a) The face setting scheme. The cube is set to be the same size with the experimental cube (200nm). The vertical black line represents the electron beam. To increase the accuracy of calculation, the region near the cube and electron beam has a much finer mesh, which shows as a cylinder with a radius of 10nm close to the cube. (b) Simulated energy loss spectra. As the scheme at the right bottom shows, the blue curve stands for the calculated spectra when the beam is placed near the edge, and red curve is for the face setting.

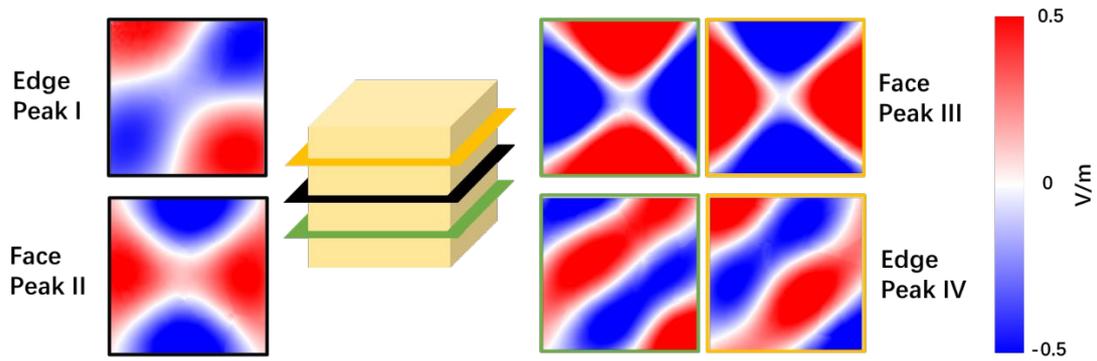

**Figure 5** Z-component (parallel to the e-beam) of electric field on x-y plane of the cube. The light-yellow cube is the simulated nano-cube. The cross section of the cube at three different height is plotted here. The yellow, black, and green cut is at the height of 50 nm (three quarters), 0 nm (half), -50 nm (quarter). On the left side is the low energy peaks (I, II) and right side is high energy peaks (III, IV). The color of figure boundary indicates the cut position. Beam is placed at left side of the square for face setting and top left for edge setting (same as **figure 4b**).